\documentclass[prd,twocolumn,letterpaper,floatfix,showpacs,amsmath]{revtex4} 
\newlength{\picwidth}

\usepackage[dvips]{graphicx}

\newcommand{\bj}[1]{\mbox{\boldmath $#1$}}

\begin{document}

\title{A Nonlinear Coupling Network to Simulate the Development of the r-mode Instablility in Neutron Stars
II. Dynamics.
}

\author{Jeandrew Brink}
\affiliation{Center for Radiophysics and Space Research,
Cornell University, Ithaca NY 14853}
\author{Saul A Teukolsky }
\affiliation{Center for Radiophysics and Space Research,
Cornell University, Ithaca NY 14853}
\author{Ira Wasserman }
\affiliation{Center for Radiophysics and Space Research,
Cornell University, Ithaca NY 14853}

\begin{abstract}
Two mechanisms for nonlinear mode saturation of the r-mode in neutron stars have been suggested: the parametric instability mechanism involving a small number of modes and the formation of a nearly continuous Kolmogorov-type cascade. Using a network of oscillators constructed from the eigenmodes of a perfect fluid incompressible star, we investigate the transition between the two regimes numerically. Our network includes the 4995 inertial modes up to $n\leq 30$ with 146,998 direct couplings to the r-mode and 1,306,999 couplings with detuning$< 0.002$ (out of a total of approximately $10^9$ possible couplings).  
 The lowest parametric instability thresholds for a range of temperatures are calculated and it is found that the r-mode becomes unstable to modes with $13<n<15$.
In the undriven, undamped, Hamiltonian version of the network the rate to achieve equipartition is found to be amplitude dependent, reminiscent of the Fermi-Pasta-Ulam problem. More realistic models driven unstable by gravitational radiation and damped by shear viscosity are explored next.  A range of damping rates, corresponding to temperatures $10^{6}$K to $10^9$K, is considered.  
Exponential growth of the r-mode is found to cease at small amplitudes $\approx 10^{-4}$. For strongly damped, low temperature models, a few modes dominate the dynamics. The behavior of the r-mode is complicated, but its amplitude is still no larger than about $10^{-4}$ on average.  For high temperature, weakly damped models the r-mode feeds energy into a sea of oscillators that achieve approximate equipartition. In this case the r-mode amplitude settles to a value for which the rate to achieve equipartition is approximately the linear instability growth rate.

\end{abstract}

\maketitle

In a previous paper \cite{JdB1} (henceforth Paper I), we presented details on the computation of the coupling coefficients and damping/driving rates for the generalized r-modes of a uniform density, uniformly rotating star in the limit of slow rotation. In this paper we explore the nonlinear dynamics of the oscillator network constructed in Paper I. 
We concentrate on the results that will have physical implications for the saturation of the r-mode instability in neutron stars. The astrophysical motivation for this problem was given in Paper I.

Consider a sea of coupled oscillators obeying the amplitude equations 
\begin{equation}\dot{c}_A(t) - iw_Ac_A +\gamma_Ac_A = -i\frac{w_A}{\epsilon_A} \sum_{BC}\kappa_{\overline{A}BC}  c_B c_C~,
\label{eq:ampeq}
\end{equation}
where the coupling coefficients $\kappa_{\overline{A}BC} $, the driving/damping rates $\gamma_A$ and the frequencies $w_A$ have been derived from the fully nonlinear problem by means of second order perturbation theory. Only the most significant nonlinear and driving terms have been retained. Thus, we assume that modes never achieve large amplitudes $c_\alpha$, and explore the dynamics in the limit of weak non-linearity. The validity of this assumption can be assessed in retrospect.

The oscillator network for the r-mode problem has many distinct properties  inherited from the original physical problem: The rescaled frequencies of the oscillators are bounded between -1 and 1,  the connectedness of the coupling graph is determined by the selection rules and the general background of the coupling coefficients provides the landscape on which the dynamics will take place.

In neutron stars  energy enters the system of coupled oscillators at a large scale via a comparatively weakly connected  r-mode.  The energy is drained out of the system by viscosity at smaller scales, from modes that are comparatively well connected and have several couplings to other small scale modes. The passage of energy from the large scale to the damping scale is what we propose to study as a mechanism for saturating the r-mode instability.

Many examples of driven fluid systems exist in nature; Rayleigh-Bernard convection \cite{RBC} and  Taylor-Couette flow \cite{CrossHohenberg} are two particularly well-studied examples. Both display a rich pattern-forming transition from laminar base flow to more turbulent regimes. The r-mode problem can be expected to display similar complexity.  The experiments that follow are designed to indicate the  types of behavior that may arise in different parameter regimes representative of those in a neutron star. 

We begin by considering the simplest coupled system of three modes in Section~\ref{ThreeMODE}. Using ideas obtained from the three mode problem we compute possible parametric instability thresholds that arise from the nonlinear coupling \cite{JdB1} of the inertial modes \cite{Bryan} of an incompressible uniformly rotating fluid. We investigate the effect of temperature on the thresholds of the inertial modes  of the three mode system. We use the analytic results for the instability thresholds and oscillation periods obtained in  Section \ref{ThreeMODE} to check our integration scheme and to indicate which amplitude regimes are important. 

The integration of approximately 5000 complex oscillators, in which all the inertial modes with $n\leq 30$ are included, coupled by approximately $1.3\times 10^7$ coupling coefficients, for integration times of $>10^7/2\Omega$  requires equation~\eqref{eq:ampeq} to be implemented as efficiently as possible.  Our approach to implementing \eqref{eq:ampeq} is detailed in Section~\ref{BuildRHS}. 

We build up our intuition about the behavior of our coupling network in Section~\ref{SmallmodeIntegrations} by integrating the triplet of modes in our network that go unstable first, namely the triplet having the lowest parametric instability threshold. A second and third direct coupling to the r-mode are then systematically added. Any simulation starting from negligibly small initial conditions will during its initial evolution display the dynamics observed for these low order systems.    

In Section \ref{SectionLargeHam} we begin our investigation of the large modal system by observing how rapidly if at all equipartition of an initial large excitation in the principal r-mode takes place.  The opposing concepts of ergodicity and phase coherence are introduced, and an indication of the nonlinear timescales is established.

Two damping regimes exist. When dissipation is strong, the daughter modes that couple directly to the r-mode can damp out the r-mode's  amplitude growth  effectively. The daughter-daughter couplings may alter the dynamics but are not essential in halting the growth. When dissipation is weak,   the modes to which the r-mode goes unstable do not  damp its growth effectively and multiple generations of daughter-daughter couplings are necessary to transfer the energy to the damping modes.  To explore this case fully we need to introduce more  couplings.  However, it is impractical to include all the couplings since their number scales roughly like $\sim n^8$. A criterion for identifying  the   important couplings is essential to render the calculation tractable. Our method of dealing with the proliferation of couplings is discussed in Sections~\ref{BuildRHS} and ~\ref{detuningsection}.

 We conclude our investigation by observing the dynamics of a few weakly damped systems in the high temperature regime. In this regime the transition to more Kolmogorov-like spectra is made and a large number of modes are excited. 

\section{Three Mode Systems}
\label{ThreeMODE}
The smallest non-linear system consists of three modes and one coupling. 
The undamped Hamiltonian system yields an analytic solution in terms of elliptic integrals \cite{CarlsonCubic} as shown in  
Appendix~\ref{AppendThreeHams}. 
The period of the amplitude oscillations  depends on the initial conditions of the system 
and can be accurately calculated numerically \cite{NR}. For the case of a large  amplitude parent coupling to small amplitude daughters the period increases as the parent amplitude decreases. It was this dependence on initial amplitudes of the nonlinear oscillation frequency that was used  \cite{JdB3} to explain the timescales involved in the catastrophic decay of the ``pumped up'' state of the r-mode observed in hydrodynamical  simulations \cite{Ster2}.

There exists a minimum parent amplitude below which no oscillations in amplitude occur. This amplitude is the no-damping limit of the  parametric instability threshold discussed in the next paragraph.  Above this threshold very long oscillation periods are realized.  Accurately following these oscillations, which take place over thousands of periods of the oscillators involved, is a challenge to our numerical code, and was one of the tests used to verify the reliability of our integration scheme.

Study of the damped system introduces the concept of the parametric instability, which sets the minimum amplitude  the r-mode  has to reach before it can excite other modes in the system. No energy is transfered out of the driven mode until the instability threshold is reached. 
The phase space of the damped three mode system was thoroughly explored by Dimant \cite{Dimant}  and Wersinger et. al \cite{Ott}. The stationary solutions and parametric instability threshholds are compactly re-derived in Appendix~\ref{parametricAPPENDIX}. 
The parametric instability threshold  for an unstable parent $\alpha$ sharing excitation with two daughters $\beta$ and $\gamma$ is
 \begin{equation}|c_\alpha|^2 = \frac{\gamma_\beta \gamma_\gamma}{4w_\gamma w_\beta \kappa^2} \left[1+\left(\frac{\delta w}{\gamma_\gamma+\gamma_\beta}\right)^2\right]~.
\label{ParINSTB2}
\end{equation}
Here $\gamma_\beta$, $\gamma_\gamma$ are the damping rates and $w_\beta$, $w_\gamma$ are the frequencies of the daughter modes, while $\delta w = w_\alpha - w_\beta - w_\gamma$ is the detuning.  Note that the instability threshold does not depend on the linear growth rate of the parent and is subject to the correct relationship between the signs of the mode frequencies (some triplets never go unstable as shown in Appendix~\ref{parametricAPPENDIX}).

\begin{figure*}
\includegraphics[width=\picwidth]{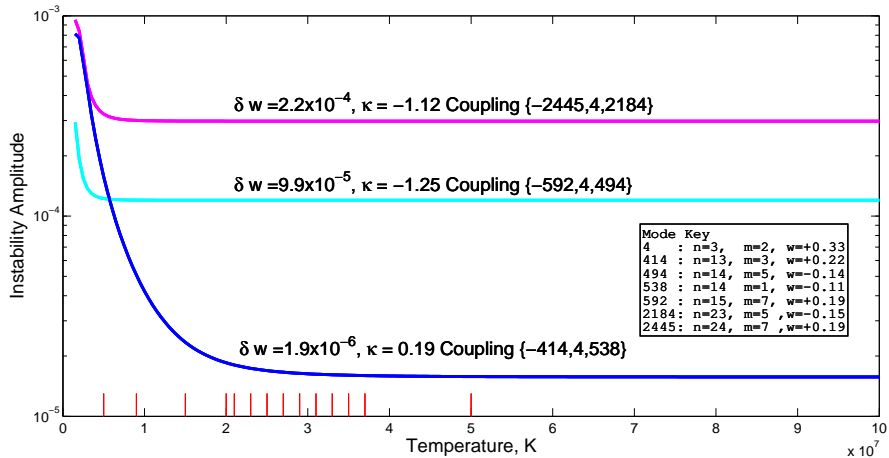}
\caption{Temperature dependence of lowest 3 parametric instability thresholds of the r-mode to the inertial modes.
} \label{ParametricInstThres}
\end{figure*}

The parametric instability thresholds for the 146,998 direct couplings to the r-mode were calculated for a range of temperatures.  All modes with $n\leq 30$ were included. The behavior of the lowest three thresholds as a function of temperature is shown in Figure~\ref{ParametricInstThres}

The graph begins with the temperature at which gravitational radiation first destabilizes the r-mode of a neutron star rotating at $\Omega = 0.37\sqrt{\pi G \rho}$, namely where the balance between the damping due to shear viscosity of the r-mode and the GR driving rate occurs. At this temperature all modes have strong viscous dissipation.   The r-mode has to attain a relatively large amplitude before the nonlinear term dominates the linear damping of the daughters (equation~\eqref{ParINSTB2}). By contrast in a system where the daughters are less strongly damped  parametric instability occurs more readily.  As the damping is decreased the instability amplitudes decrease monotonically until the limiting value of $|\delta w/4\kappa \sqrt{ w_\beta w_\gamma}|$ is reached.

The inset in the graph gives the conversion of the numbers labeling a particular coupling to the eigenmodes labeled according to degree, $n$, and order, $m$, of the Legendre functions that make up the eigenmodes' functional representation. The lowest parametric instability in the network we considered occurred for modes with reasonably high degree, namely $n=13$ and $n=14$. For low enough temperature, the shear damping is large  enough so that the three mode system can dissipate all the energy with which the r-mode is driven, stopping its growth.  However, as the damping decreases the r-mode is driven at a rate that cannot be halted by mere three mode coupling, and more modes become important.

\section{Numerical Implementation of the Amplitude Equations}
\label{BuildRHS}

We implement the amplitude equations~\eqref{eq:ampeq} numerically by means of a Bulirsch-Stoer method with adaptive stepsize control \cite{NR}. 
Adaptive stepsize is a feature that is essential to any scheme adopted. The evolution begins with slow exponential growth described by linear perturbation theory, during which there is little excitation of the rest of the network, allowing the use of larger stepsizes, but proceeds to the excitation of many modes via nonlinear effects, which demands shorter stepsizes. 
 A further criterion for the selection of an integrator is that it must minimize the number of function evaluations of the right hand side (RHS), since  looping over a large number of coupling coefficients becomes very expensive.   

Another  integration scheme that was explored was the fourth-order Runge-Kutta scheme, also with adaptive stepsizing \cite{NR}. 
However, this method  displayed a larger systematic secular drift in the energy of a Hamiltonian system than the Bulirsch-Stoer method did.  Symplectic integrators were also investigated, but we were not able to find one with adaptive stepsizing that could compete with the high order accuracy of the Bulirsch-Stoer method with roughly the same number of function evaluations.  Furthermore, as soon as dissipation is added the advantage of using the symplectic integrator is no longer clear. Semi-implicit extrapolation methods, such as the {\it stifbs} routine \cite{NR}, allow larger stepsizes to be taken at the cost of the construction of the Jacobian matrix. Although equation \eqref{eq:ampeq} is well suited to the construction of the Jacobian analytically, limitations on memory, and the cost of communicating large matrices in a parallel architecture, made this integrator viable only for a system with a small number of modes. Regardless of the integrator used, the majority of the CPU time will be spent in the evaluation of the RHS of equation~\eqref{eq:ampeq}. This is a major limiting factor in integrating over very long times.

To implement the RHS of equation\eqref{eq:ampeq} we maintain an ordered list of all possible nonzero couplings $\kappa_{\overline{A}BC}$. For each entry in the list we store the three modes that are coupled, their coupling coefficient, and the frequency detuning.  The modes are identified by the one dimensional index $j$ introduced in Paper I which arranges the inertial modes labeled by the three numbers ($n$,$m$,$k$) consecutively. A vector $\bj{c}$ contains the complex amplitudes indexed by the same index $j$. 

After initializing the derivative terms to zero ($\dot{\bj{c}}=0$), we loop through all the couplings, adding the contribution from each nonlinear three mode coupling to the RHS. 
Since the amplitude vector is indexed by the mode number this can be done very efficiently. (Note that all the coupling coefficients for the problem we are considering are real.) 

Two possible cases exist. If $\beta = \gamma$, 
then the rule for updating the amplitudes due to this coupling is 
\begin{eqnarray}
\dot{c}_\alpha &\rightarrow&\dot{c}_\alpha - iw_\alpha \kappa_{\overline{\alpha}\beta\beta} c_\beta c_\beta \nonumber\\
\dot{c}_\beta &\rightarrow &\dot{c}_\beta - i2w_\beta \kappa_{\overline{\alpha}\beta\beta} c_\alpha c_\beta^* ~.
\end{eqnarray}
Otherwise (the general case), when the coupling is among three distinct modes, the rule for updating the RHS is
\begin{eqnarray}
\dot{c}_\alpha &\rightarrow& \dot{c}_\alpha - i2w_\alpha \kappa_{\overline{\alpha}\beta\gamma} c_\beta c_\gamma \nonumber\\
\dot{c}_\beta & \rightarrow& \dot{c}_\beta - i2w_\beta \kappa_{\overline{\alpha}\beta\gamma} c_\alpha c_\gamma^* \nonumber\\
\dot{c}_\gamma &\rightarrow & \dot{c}_\gamma - i2w_\gamma \kappa_{\overline{\alpha}\beta\gamma} c_\alpha c_\beta^* ~. 
\end{eqnarray}
The factor of 2 takes into account the symmetry in the last 2 components of the coupling. (See \cite{Katrin1}, equation (4.17).  Note that for our modes, the m-selection rule forbids the last term of equation (4.17) from occuring.)

Constructing the RHS in this manner lends itself to parallelization especially when systems with large numbers of coupling coefficients are integrated.   Distributing the coupling coefficients over the various processors,  letting each processor compute the contribution due to its local set of coupling coefficients, and then combining the result by communicating only the relatively small derivative vector results in a very efficient parallel code, with only a small overhead in communication costs. The computation time decreases roughly linearly with the number of processors added for systems with a large number of coupling coefficients. 

After communication between the processors has taken place, the linear part of the equation is added using\begin{equation}\dot{c}_\alpha \rightarrow \dot{c}_\alpha+(i w_\alpha-\gamma_\alpha) c_\alpha.\end{equation}
This implementation follows each individual oscillation of the real and imaginary parts of a modes involved.  It yields very accurate results for energy  conservation of the Hamiltonian system, 
\begin{equation}
E = \sum_\alpha c_\alpha c_\alpha^* - 4\sum_{\kappa_{\overline{\alpha}\beta\gamma}}^{\beta\neq\gamma} \kappa_{\overline{\alpha}\beta\gamma} \Re (c_\alpha^*c_\beta c_\gamma) -2\sum_{\kappa_{\overline{\alpha}\beta\beta}}\kappa_{\overline{\alpha}\beta\beta}\Re(c_\alpha^*c_\beta ^2)~.
\label{eq:ENERGY}\end{equation} 
In equation \eqref{eq:ENERGY} summation is implied over all  nonzero couplings, (not mode indices as used in \cite{Katrin1}) and $\Re()$ indicates the real part of a complex number.  

For the scheme outlined above, the accurate integration of the frequency terms $i w_\alpha c_\alpha$ limits the timesteps to less than a rotation period, making it impractical for achieving long integration  times of approximately $10^7/2\Omega$.
The coordinate transformation 
\begin{equation}
c_\alpha = C_\alpha e^{iw_\alpha t}
\end{equation}
removes the explicit oscillatory term.  However, it does so at the expense of introducing a few rapidly oscillating  terms, proportional to $e^{i\delta w t}$ (see equation~\eqref{eiwt3mode}) in the nonlinear coupling terms. These terms may have frequencies up to three times those found in the original system, equations \eqref{eq:ampeq}. Thus the coordinate transformation itself results in no significant speed up of the integration.  However, the couplings that play the most important role in the long term evolution of the problem are those near resonance, with detuning close to zero.  Couplings with large detuning that oscillate very rapidly limit the step size of the simulation, but on average contribute very little to the longer term dynamics. Retaining only couplings with a detuning less  than some pre-selected cutoff allows much larger timesteps to be achieved. For example retaining only couplings with $\delta w < 0.002$ for the $n\leq 30$ system allows stepsizes  $\approx 400/2\Omega$ to be taken initially and as more modes become active this decreases to about $40/2\Omega$ . 
The validity of this assumption and the effect of choosing different cutoffs are explored in greater detail in Section \ref{detuningsection}.

Imposing an upper limit on $\delta w$ also  decreases the number of couplings retained in the nonlinear interaction terms, thus speeding up the calculation. Additional speedup is achieved by dynamically adding  couplings as the simulation progresses. Consider an initial condition in which all the modes lie below a certain noise threshold,  chosen small enough that the non-linear interactions are negligible.  Since at most a few r-modes are unstable at first, the evolution is simply exponential growth of these modes. As these unstable modes grow in amplitude, they drive the modes that couple to them directly.  All couplings that do not couple directly to the unstable modes during this early phase can be neglected. As an unstable mode passes its first parametric instability threshold, the daughter modes coupling directly to this mode will increase in amplitude, rise above the noise, and start playing a significant role in the dynamics of the simulation. When the daughters rise above a preselected threshold their contribution to the nonlinear driving will become important for other modes, and all direct couplings to the daughters are then added to the existing coupling list. The step size is decreased and the integration continued with the updated coupling list. In practice, to prevent repeatedly restarting the simulation with new couplings, when a certain mode is tagged as becoming unstable, the direct couplings to all latent modes with an amplitude  within $50\%$ of the amplitude of the unstable mode are also activated. Successively adding couplings in this manner allows the simulation to move through the initial phase of computation more rapidly.

We used a number of checks on  our integration scheme.  Analytic results for the three mode Hamiltonian system are reproduced, including the sensitive dependence of the nonlinear period for parent amplitudes just above the parametric instability threshold. The parametric instability thresholds for the damped three mode system are accurately observed. (These thresholds continue to be used as an analytic guide for larger systems although there is no strict mathematical proof that ensures the modes should become unstable at the lowest three mode parametric instability.) We also checked that for zero damping and growth rates the energy of the Hamiltonian system \eqref{eq:ENERGY} drifts very slowly compared to the growth rate of the r-mode instability.

\section{Integrations of Small Systems}
\label{SmallmodeIntegrations}
\begin{figure*}[htb]
\includegraphics[width=\picwidth]{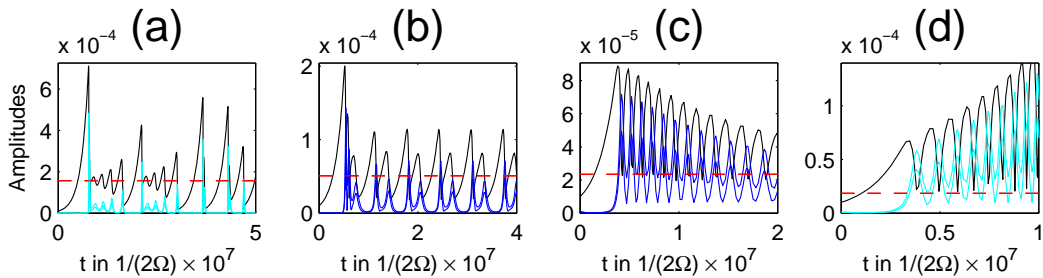}
\caption{Numerical solutions to the three mode problem with the lowest ($\gamma_{\beta},\gamma_{\gamma}\rightarrow 0$) parametric instability threshold  for a variety of temperatures (a) $T = 5\times 10^6 K$ (Chaotic motion), (b) $T = 9\times 10^6 K$ (two point periodic cycle), (c) $T = 1.5\times 10^7 K$ (approaches one point periodic cycle) and (d) $T = 2\times 10^7 K$ (diverging solution). The actual ( $\gamma_{\beta}, \ \gamma_\gamma\neq 0$ ) parametric instability thresholds are indicated with dashed lines.
} 
\label{ThreeModSol} 
\end{figure*}
As was mentioned in Section~\ref{ThreeMODE} and Appendix~\ref{parametricAPPENDIX}, the damped three mode system has been studied extensively  \cite{Dimant} \cite{Ott}\cite{Dziem}. However, there have been only a few attempts at quantifying the dynamics in the case where the steady state solution is unstable,  for example, when the detuning is small or the daughter mode damping is slow. As a result it is instructive to have a closer look at the particular three-mode coupling in our network that goes unstable first for various damping rates, as its  dynamics sets the initial development of any simulation. The existence and character of steady state solutions and the behavior of  the three mode problem when the daughters can no longer effectively damp out the parent govern  which  modes are excited subsequently.

The evolution of the three-mode system with the lowest (zero damping) parametric instability threshold is shown in Figure~\ref{ThreeModSol} for various damping rates relevant  to our problem.

The qualitative picture in which the parent mode is damped by two daughter modes and reaches a steady state solution presented by Arras et al \cite{Saturation} is somewhat simplistic. Rather, the parametric instability threshold's strong dependence on frequency detuning (see equation \eqref{ParINSTB2}) implies that a coupling with small detuning will go unstable first.  Even if this coupling is strongly damped, the small detuning places it in a regime (as clearly depicted by Dimant in Figure 1(b) of his paper \cite{Dimant}) where no constant amplitude states exist, and all solutions vary with time. Furthermore, the closer to resonance the coupling is, the more complex the solutions become,  displaying increasingly many periods and entering chaotic regimes \cite{Ott}. The first unstable coupling of the $n=3$, $m=2$ r-mode falls in this regime, so large unpredictable changes in its amplitude may result, perhaps causing the r-mode to be hard to detect conclusively in the strong damping regime.    

In the plots (a) to (c) of Figure~\ref{ThreeModSol} the daughter modes are sufficiently damped to saturate the growing r-mode via a single three-mode interaction.  The maximum amplitude the r-mode reaches is roughly of the same order of magnitude as the parametric instability threshold. As shown in Figure~\ref{ParametricInstThres}, large damping  has the counter-intuitive effect of resulting in large amplitudes. One of the identifying features of these strongly damped systems is the saw-tooth shape of the modal evolution: exponential growth of the parent,  followed by excitation of the daughter modes, and rapid dissipation.  This saw-tooth feature is also seen in evolutions with large numbers of modes.  As the damping is decreased, the nature of the three mode solution changes from a system bounded in phase space and energy to a system that is diverging with ever increasing frequency oscillations, Figure~\ref{ThreeModSol}(d). High temperature neutron stars most likely fall into this regime. 

\begin{figure*}[htb]
\includegraphics[width=\picwidth]{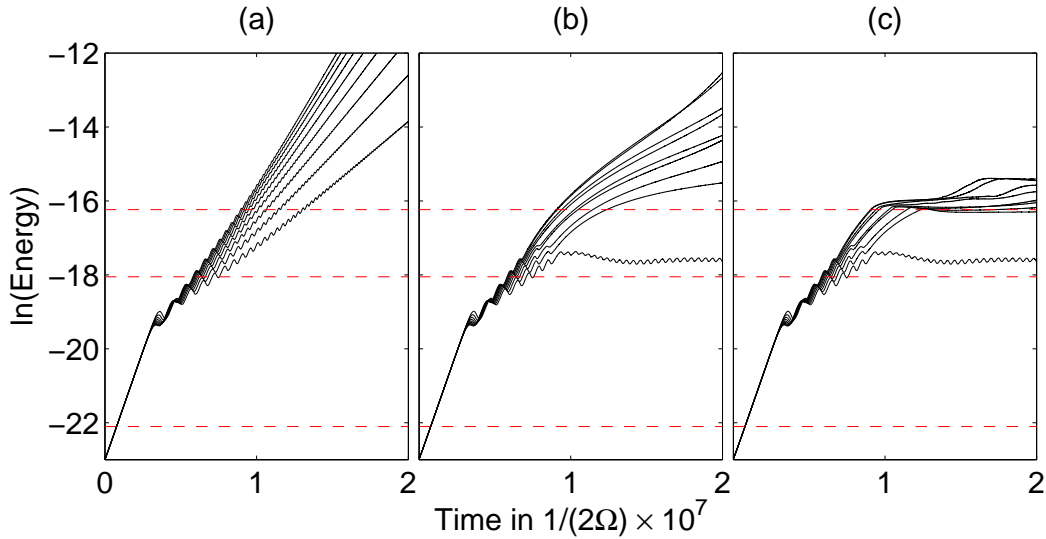}
\caption{Evolution of the Hamiltonian energy for temperatures in the range $T = 2.1\times 10^7$K to $T = 3.7\times 10^7$K in increments of  $2\times 10^6$K. Plot (a) shows the diverging three mode system with the lowest parametric instability. Plot (b) shows the system used in (a) with the coupling that yields the second lowest parametric instability threshold added. Plot (c) shows the case with the three lowest parametric instability thresholds included. Dashed horizontal lines indicate the linear energy corresponding to the parametric instability thresholds. 
} 
\label{GrowthRates} 
\end{figure*}

The energy for diverging three mode systems is shown in Figure~\ref{GrowthRates}(a) for various temperatures. A period of exponential growth increases the parent amplitude above the parametric instability threshold, whereupon energy begins to be transfered to the daughter modes. Subsequently the total energy diverges at a slower rate. The divergence rate increases roughly linearly with decreasing daughter mode damping. High temperature simulations diverge more quickly than the more damped low temperature simulations.   
Introducing the couplings with the  second and third lowest parametric instability  thresholds decreases the growth rate of the Hamiltonian energy further, as is shown in  Figures~\ref{GrowthRates}(b) and~\ref{GrowthRates}(c).  
Empirically, the dynamics begin to differ from the unstable three mode evolution once the parametric instability thresholds of the additional mode couplings are exceeded. The instability thresholds of the couplings used to make these plots as well as the modes involved are shown in Figure~\ref{ParametricInstThres}.  The first and second parametric instabilities occur for relatively low order modes with $n\sim 13 ,14, 15$ where the viscous damping cannot stop the growth entirely.  The third instability is to modes with $n\sim 23, 24$, where dissipation is large enough to stop the growth, so the system energy levels off. 

\begin{figure*}[htb]
\includegraphics[width=\picwidth]{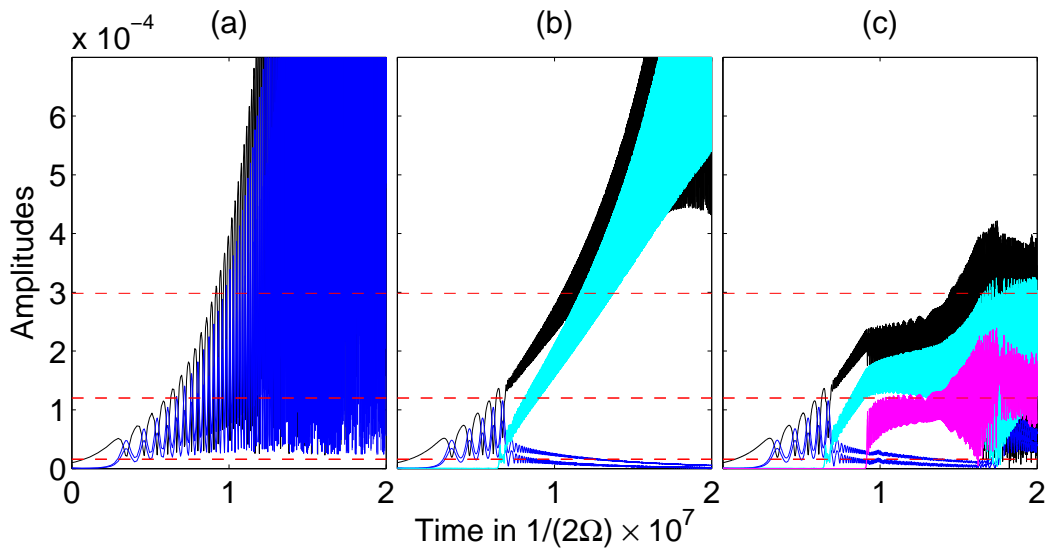}
\caption{(Color online) Evolution of the modal amplitudes for the systems discussed in Figure~\ref{GrowthRates} for the case with  $T = 3.7\times 10^7$K Dashed lines indicate the parametric instability amplitudes. The n=3, m=2 r-mode is indicated in black, other colors are chosen to match the color coding for the instability thresholds used in Figure~\ref{ParametricInstThres}.
} 
\label{ThreeMGrowthRates} 
\end{figure*}

For the most rapidly diverging,  $T=3.7 \times 10^7$K case shown in Figure~\ref{ThreeMGrowthRates}(a),  the isolated three mode solution has wild oscillations with ever increasing frequency and amplitudes. Adding the second coupling Figure~\ref{ThreeMGrowthRates}(b) changes the dynamics dramatically: After the second parametric instability is exceeded, the second coupling dominates,  while the initially unstable three mode coupling is almost entirely suppressed.  The third coupling Figure~\ref{ThreeMGrowthRates}(c) becomes important at an amplitude between the parametric instability thresholds for the second coupling  and the third couplings. When the r-mode reaches the third parametric threshold, the dynamics change once again; thereafter, it appears as if the second coupling no longer suppresses the first triplet, whose daughter modes begin to grow again.
 
General three mode systems  have been extensively study by Baesens et al \cite{Baesens} and can display very complicated behaviour in various regions of parameter space. Adding additional oscillators further increases the phase space to be explored, and the complexity of solutions that can be attained. Our study of these systems  is not comprehensive, and only serves as one example of what can happen as additional degrees of freedom are added.  It is, however, the example that is most relevant to our larger network. One can hope to recognize  the effects of individual three mode couplings in the larger systems. This can allow one to differentiate between the effects of individual three mode couplings and those  that result from the combined couplings of many modes.

The previous examples do not take into account the couplings among the daughter modes which will be present in a larger system. As we turn to  coupled systems with numerous modes, additional factors such as the effective number of degrees of freedom in the phase space available at a certain time, and the structure of the coupling network, will become important. 

\section{Larger Systems}
\label{LargeIntegrations}
During subsequent sections we explore the characteristics of an oscillator network that includes all of the inertial modes up to $n=30$, with the gravitational driving fixed to the value for a star rotating at $\Omega = 0.37\sqrt{\pi G \rho}$ (i.e. $\nu=695$ Hz). This corresponds to a principal ($n=3$, $m=2$ or $j=4$) r-mode driving rate of $\gamma_4 = 5.9\times 10^{-7} (2\Omega)$ in our dimensionless units.  Including all  couplings among these modes is impractical. There are a total of approximately $10^9$ nonzero couplings. So we only account for  couplings among modes with frequency detunings below a pre-selected maximum value.  The effects of employing different detuning criteria are explored in greater detail in Section~\ref{detuningsection}. The majority of simulations are integrated using a criterion of $\delta w < 0.002$ which results in the inclusion of 1306999 couplings between the 4995 modes.  We begin by integrating the Hamiltonian system; this checks our integration scheme, and also yields insights on the interaction among numerous modes. We then continue to explore our large network for various shear damping regimes in Sections~\ref{Dampedsystems}.

\section{Large Hamiltonian Systems}
\label{SectionLargeHam}
\begin{figure}[h]
\includegraphics[width=\columnwidth]{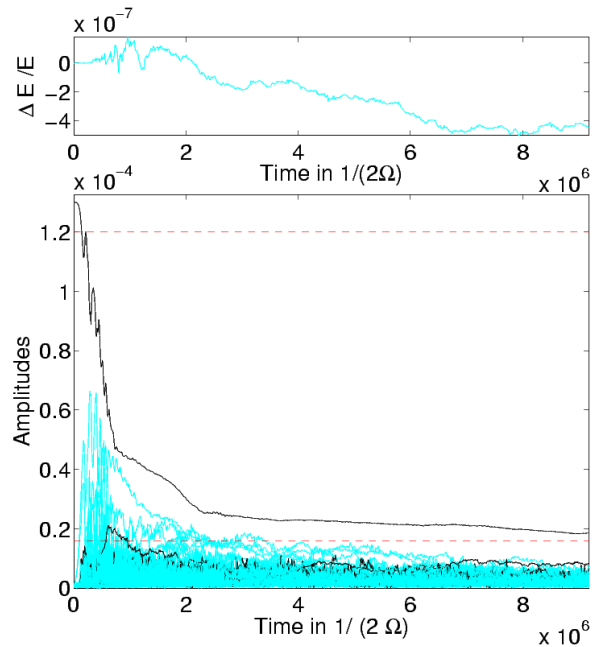}
\caption{(Color online) Evolution of the Hamiltonian system including the 1306999 couplings with $\delta w < 0.002$. An initial excitation of $1.3\times 10^{-4}$ is placed in the principal r-mode. $n=m+1$ r-modes are indicated in black, all other modes are indicated in light grey/cyan on the lower panel.  The parametric instability thresholds are indicated with dashed lines. The relative energy change durning the simulation is indicated on top. } 
\label{Hamiltonian} 
\end{figure}

Without damping and driving, the Hamiltonian energy of the oscillator network (equation~\eqref{eq:ENERGY}) should be conserved for any subset of couplings. Having an integration scheme that accurately conserves energy over a long period of time is particularly important because otherwise the slow growth and damping rates of the r-mode problem may be overshadowed by spurious numerical effects.  Figure~\ref{Hamiltonian} (top) shows the fractional change of the Hamiltonian energy ($E$) for a system in which the r-mode is initially excited to an amplitude just above the second parametric instability threshold. Note that initially $E=1.69\times 10^{-8}$.
The fractional energy change was $4\times 10^{-7}$ over a time period of $9\times 10^6/2\Omega$; in the same time span, exponential growth of the r-mode would result in $\Delta E / E \sim 10$. Although the growth rates in a large nonlinear system  may be much smaller than in linear theory, we feel confident that our simulations can accurately track $\Delta E /E \geq 10^{-5}$ on timescales $\sim 10^7/2\Omega$ for fully nonlinear systems including the gravitational radiation instability and viscous decay.

A large Hamiltonian system involving three mode couplings that has been extensively studied is the Fermi-Pasta-Ulam problem (FPU) \cite{FPU1} \cite{FPU2}. First performed in 1954, this numerical study of a chain of coupled nonlinear oscillators yielded a number of surprises. Fermi et al. expected to observe relaxation to an equipartitioned state, when starting from an initial state in which only a large scale mode was excited.  Instead, they observed that only a few modes were excited; moreover after a long enough time, the system appeared to return to approximately the initial state (recurrence). The remaining modes never rose out of the noise. Subsequently the FPU problem has been studied extensively \cite{Ford}\cite{Lich1}\cite{Lich2}\cite{Lich3}. Because the dispersion relation of the FPU problem admits very few direct resonances among modes, in retrospect one might have expected equipartition to fail to be achieved for the small initial amplitudes chosen by Fermi et. al.  Although the FPU problem does tend to equipartition  for initial amplitudes above a certain amplitude threshold \cite{FPUthreshholds}, its failure to do so at small amplitudes  cautions us that a Kolmogorov spectrum \cite{Zak} (of which the equipartition solution is a special case), may never arise in a system in which relatively little energy is fed into a mode that has few direct resonances to other modes.

In the network we are considering the r-mode has very few near resonances. The nearest resonance, with a detuning of $1.9\times 10^{-6}$, is associated with the coupling that has the smallest parametric instability threshold in Figure~\ref{ParametricInstThres}.  The small scale daughters to which the r-mode goes unstable first are directly coupled to many more modes.  The dispersion relation for inertial modes, which implies that at a given $n$ there are at least $n$ different modal frequencies roughly uniformly distributed between $-2\Omega$ and $2\Omega$, ensures that small scale modes may have many near-resonant couplings. After the initial bottleneck to energy flow out of the large scale $n=3$, $m=2$  r-mode, equipartition among the daughter modes is much more probable than  in the FPU system. The lower panel of Figure~\ref{Hamiltonian} displays the decay of an r-mode with initial amplitude just larger than the second parametric instability threshold to a more ``equipartitioned state''.  At late times, the r-mode amplitude appears to approach the first parametric instability threshold, rather than a true equipartitioned state. However, we did not run our code long enough to verify that the r-mode amplitude never falls below the first instability threshold.

\begin{figure}[h]
\includegraphics[width=\columnwidth]{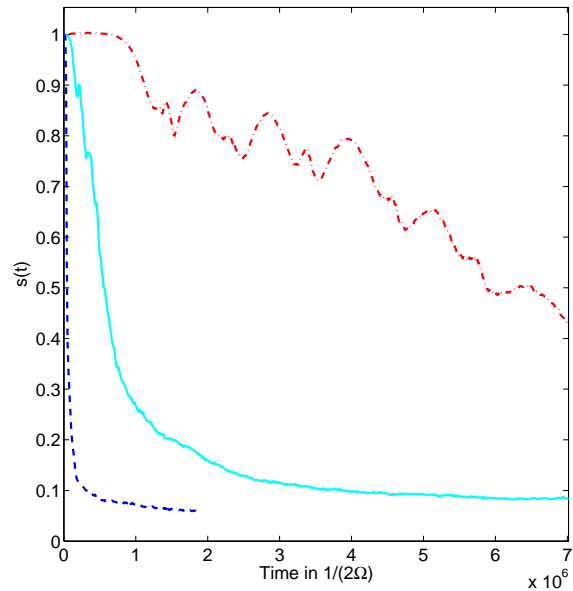}
\caption{(Color online) Approach to equipartition of the Hamiltonian system. The time evolution of the spectral entropy
for various initial r-mode amplitudes : The dashed line is $c_\alpha(0) = 5\times10^{-4}$(blue), the solid line is $c_\alpha(0) = 1.3\times10^{-4}$   (cyan) and the dot-dashed line is $c_\alpha(0) = 0.8\times10^{-4}$(red).
} 
\label{HamEntrop} 
\end{figure}

A measure of the degree of equipartition of our system as a function of time is  the spectral entropy \cite{FPUthreshholds2}
\begin{equation}
S(t) = -\sum_j w_j(t)\ln\ w_j(t) 
\label{entropyeq}
\end{equation}
where the weights $w_j$ are given by the fraction of linear energy in the $j$th mode: \begin{equation} w_j = \frac{|c_j|^2(t)}{\sum_j |c_j|^2(t)}~.\end{equation}  
In equilibrium  $S(t)$ reaches a maximum value of $S_{max}= \ln(N_{modes})$ where $N_{modes}$ is the total number of modes coupling. In this state the system is likely to become ergodic and could be described by the random phase approximation. In order to compare the evolutions of two different systems, the entropy can be normalized using
\begin{equation} s(t) = \frac{S(t)-S_{max}}{S(0)-S_{max}} \end{equation}
where $s\approx 1$ when the energy is restricted to a small number of modes, and $s = 0$ for equipartition. Figure~\ref{HamEntrop} shows $s(t)$ for three different initial r-mode amplitudes, chosen to lie between successive parametric instability thresholds. 
For larger initial energies, the decay to an equipartitioned state occurs rapidly. As the initial amplitude is decreased, the timescale to reach equipartition increases. A system with driving and damping may remain far from equilibrium, with dynamics dominated by just a few modes if the time to equipartition exceeds the damping and driving timescales.   In this situation, the random phase approximation fails, and the evolution of the system may be rich and varied.  The pattern-forming transition to turbulence, which is well-studied in Rayleigh-Bernard convection, is an example of this regime \cite{RBC}.

\begin{figure}[h]
\includegraphics[width=\columnwidth]{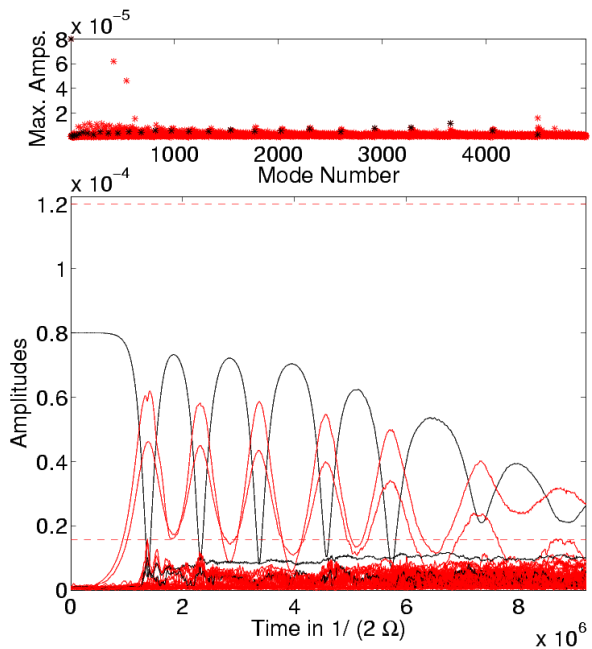}
\caption{(Color online) Modal dynamics for a Hamiltonian system with an initial amplitude $c_\alpha(0) = 0.8\times10^{-4}$ (bottom). The $n=m+1$ r-modes are indicated in black. The distribution of the maximum modal amplitude obtained with respect to mode number is shown on top.} 
\label{FPUlikeHam} 
\end{figure}

For our oscillator network, ergodicity  begins to fail for initial r-mode amplitudes between   $c_\alpha(0) = 0.8\times10^{-4}$ and  $c_\alpha(0) = 1.3\times10^{-4}$. The approach to  equipartition for  $c_\alpha(0) = 0.8\times10^{-4}$ is very slow. As is shown in Figure~\ref{FPUlikeHam},  a small subset of modes maintain apparently phase-coherent dynamics, reminiscent of the FPU problem. 

An alternative  mechanism for damping the r-mode growth is  nonlinear interactions with a sea of coupling modes. The observed decay toward equipartition can be considered as an effective many-mode damping effect.   For sufficiently large amplitudes,  details of the network appear to be insignificant and the time to equilibration rapid. In the absence of dissipation, the energy originally  in the r-mode will be shared rapidly among the network of coupled modes. 
In a very large system of modes, the ``specific heat'' is roughly
proportional to the number of modes. As a result, the amount of
energy needed to raise the effective amplitude of the coupled
system is very large. In a system that has achieved equipartition,
the amplitudes of all modes are (statistically) the same so the
energy cost to raise the amplitude of any single mode -- including
the $n=3$, $m=2$ r-mode -- by a specific amount is increased by the
number of modes in nonlinear contact.

 As the initial r-mode amplitude and  energy in the system are decreased, the details of the network become more significant, governing which modes can be excited and what pathways the energy transfer follows.  The time to equipartition becomes larger as the amplitudes decrease, effectively trapping the energy in a few modes. 
If the r-mode is driven by gravitational radiation starting from
very small amplitudes, its energy will change rapidly at first
compared with the timescale to equipartition. However,
as its amplitude rises further, the r-mode will begin to share
energy with other modes, so that its amplitude 
will increase much more slowly subsequently. Thus, the effective
rate of growth of the r-mode instability will vary as the mode amplitude
evolves, but eventually settles down to a rate slower than 
indicated by linear theory as a result of nonlinear interactions.

In Section \ref{SmallmodeIntegrations} the concept of decreasing the r-mode growth rate due to viscous damping of one or several daughter modes was introduced.
In a many-mode system, the effect of dissipation is more complex than just  a drain of energy. Large damping increases  the parametric thresholds, thus  limiting  the pathways available for energy flow, and effectively  decreasing the ``specific heat'' or many-mode damping effect, which results in a greater growth rate of the r-mode.

From an observational point of view, if an equipartitioned  solution or a more general Kolmogorov solution is achieved, gravitational radiation would be emitted over a continuous spectrum from 
the sea of inertial modes, rather than a discrete spectrum at the r-mode frequency.

\section{Large Non-Conservative Systems}
\label{Dampedsystems}
\subsection{Strong Damping Case}
\begin{figure*}[htb]
\includegraphics[width=\picwidth]{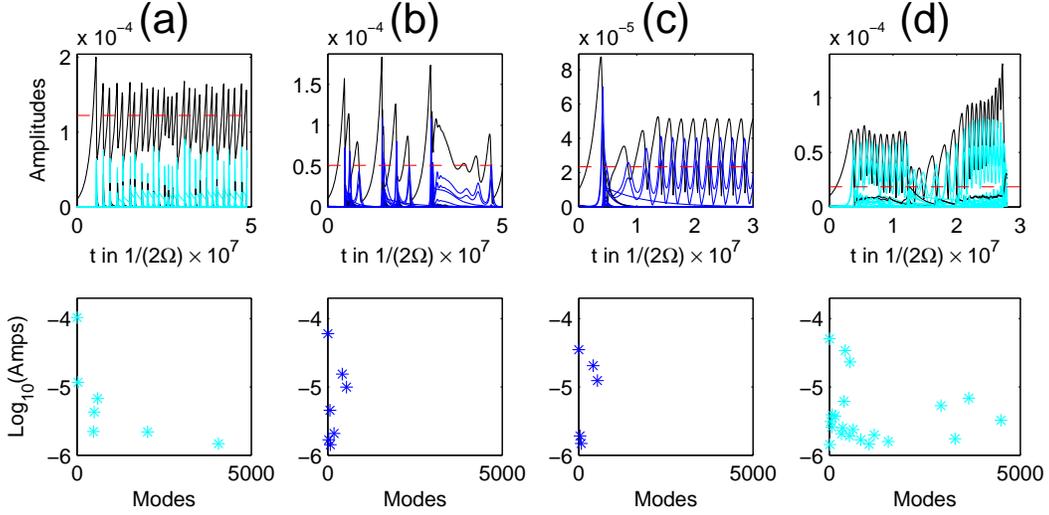}
\caption{(Color online) Top panels: Time evolution of large strongly damped systems  for temperatures (a) $T = 5\times 10^6 K$
 (b) $T = 9\times 10^6 K$  (c) $T = 1.5\times 10^7 K$  and (d) $T = 2\times 10^7 K$.   Bottom panels: Corresponding distribution of averaged modal amplitudes that rise above the initial noise level. In the top panels, the parametric instability thresholds are indicated with dashed lines.} 
\label{EvolStrongDamp} 
\end{figure*}

The inclusion  of driving and damping terms adds new nondynamical  timescales to the problem, and changes the instability thresholds. An estimate of the time scale introduced by gravitational radiation for our model is \mbox{$1/\gamma_4 = 1.7\times 10^6 /(2\Omega)$}  .  For very strong  damping, the instability threshold of the daughter modes increases, so the r-mode reaches large amplitudes before the daughters begin to grow. 
However, at these large amplitudes, the effective non-linear energy transfer rate is also rapid, resulting in rapid energy transfer from the daughters to higher order modes. These high order modes dissipate energy effectively, and the r-mode can damp almost to the noise level before the daughter modes themselves decay. At such strong damping, only a small number of  modes are important to the dynamics, and the behaviour is qualitatively very similar to the saw-tooth chaotic motion observed in the strongly damped three mode case shown in Figure~\ref{ThreeModSol}

\begin{figure}[h]
\includegraphics[width=\columnwidth]{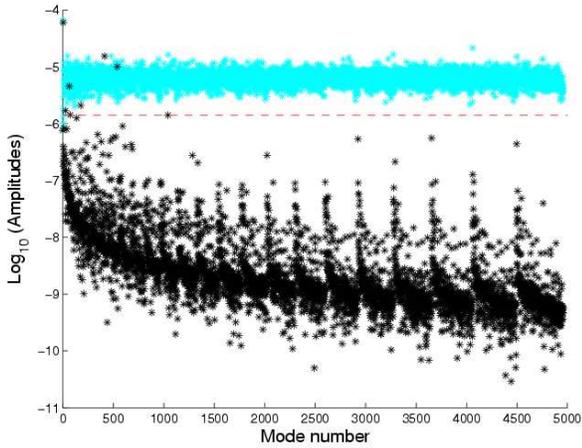}
\caption{(Color online) Modal spectra for strongly damped system, $T =9\times 10^6$K  (black) and an undamped system, ``equipartition solution'' (light/cyan). Initial noise level indicated with a dashed line.} 
\label{modalSpectrum} 
\end{figure}
The integrations of the large network of modes corresponding to the temperatures  used to produce Figures~\ref{ThreeModSol}(a) -~\ref{ThreeModSol}(d)
are shown in Figure~\ref{EvolStrongDamp}.  In these simulations only a small number of modes attain  significant amplitudes, while the rest simply decay via shear viscosity. The amplitude spectrum from the results in Figure~\ref{EvolStrongDamp}(b) is compared with the equipartitioned spectrum obtained for  zero damping with $c_\alpha(0) = 5\times 10^{-4}$ in Figure~\ref{modalSpectrum}. Both simulations start with an initial noise level of $\sqrt{2}\times 10^{-6}$ (indicated by means of a dashed line in Figure~\ref{modalSpectrum}). The imprint of the shear damping is clearly visible in the damped spectrum (compare with the analytic results in Paper I). The equipartitioned result just increases the energy in every mode by the same amount causing the spectrum to rise above the noise level, whereas there is distinct structure in the power spectrum for the dissipative system. These spectra indicate the two extreme cases that could exist in our simulations.

Note that the the set of modes to which the r-mode in Figure~\ref{EvolStrongDamp}(a) goes unstable first is different from the first unstable triplet in the Figures~\ref{EvolStrongDamp}(b)-~\ref{EvolStrongDamp}(d) and the triplet used to produce in Figure~\ref{ThreeModSol}(a). The parametric instability to which it does go unstable is correctly predicted in Figure~\ref{ParametricInstThres}.  The difference between the modes excited in \ref{EvolStrongDamp}(a)  as compared to those excited in \ref{EvolStrongDamp}(b) and \ref{EvolStrongDamp}(c) indicates the influence the first unstable triplet has on the subsequent second generation daughter modes. Figures~\ref{EvolStrongDamp}(b) and \ref{EvolStrongDamp}(c) are very similar to the pure three mode problems depicted in  Figure~\ref{ThreeModSol}(b) and \ref{ThreeModSol}(c); this is to be expected since the damping is strong enough to saturate the r-mode using the three mode coupling only. Figure~\ref{EvolStrongDamp}(d) corresponds to a diverging solution  in the three mode problem Figure~\ref{ThreeModSol}(d). This case shows a larger number of modes raising above the noise.

\subsection{The effect of detuning}
\label{detuningsection}

In our simulations, we select the couplings that we expect to play an important role in an evolution based on their detuning, because including all the couplings between the modes up to $n =30$ is impractical. Both the time to calculate all the couplings and the time to run the evolution would be absurdly long.

The suitability of a maximum detuning criterion as a method of selecting important modes can be seen from equation~\eqref{eiwt3mode}.  Modes with large detuning will oscillate very rapidly while a smaller detuning implies a slower variation and will on average have a greater effect on the modal evolution. Taken to the extreme, the selection of modes based on a detuning criterion results in including only modes that are in exact resonance.  This assumption is inherent in the random phase approximation  used by Arras et al \cite{Saturation} in their cascade solution. This simplifying assumption has been very successful in explaining spectra observed in various systems of coupling waves \cite{Zak}. The frequency spectrum for large scale modes is, however, rather sparse and all couplings to large modes have a finite detuning. The resonance assumption thus fails in the case of large scale modes as it will exclude all direct couplings to the driven modes. A finite detuning cutoff thus has to be selected.

In systems where only a few modes are excited, such as the strongly damped cases, the parametric instability thresholds, equation~\eqref{ParINSTB}, play an important role in determining the dynamics of the evolution. These thresholds are strongly dependent on the detuning,  further emphasizing the importance of near resonant couplings. In computing the parametric instability thresholds for all the direct couplings to the r-mode we found that the lowest three thresholds all had detuning $<3\times 10^{-4}$.

\begin{figure}[h]
\includegraphics[width=\columnwidth]{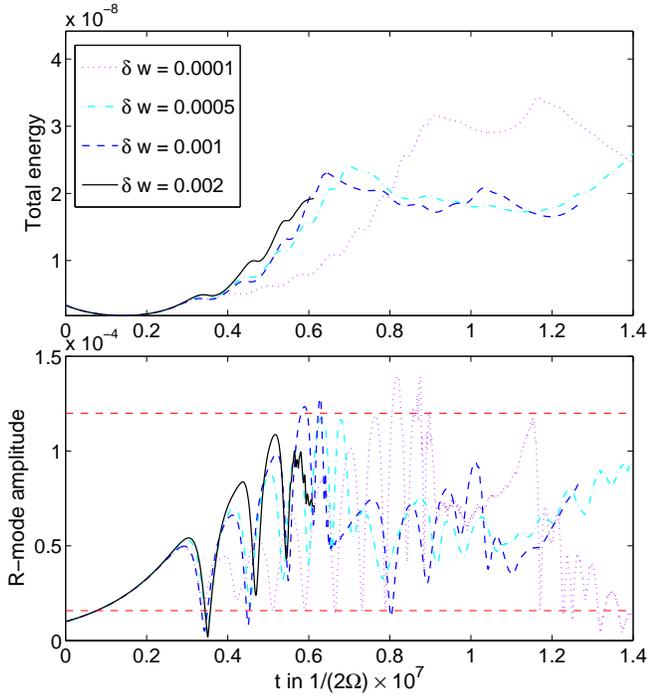}
\caption{(Color online) Evolution of the ($n=3$,$m=2$) r-mode (bottom) and total energy (top) for a $T = 5\times 10^7$K system with different detuning criteria. The parametric instability thresholds are indicated by horizontal dashed lines.} 
\label{dtComp} 
\end{figure}

To explore the effect of selecting couplings by means of a maximum detuning we ran  the coupling network  for  $T = 5\times 10^7$K system with various cutoffs in $\delta w$. The results
are displayed in Figure~\ref{dtComp} for $\delta w < 0.0001, 0.0005, 0.001, 0.002$.  
The smaller the maximum allowed $\delta w$,  the smaller the number of pathways available for the energy to leave the r-mode or subsequently excited daughter modes. A detuning criterion of $\delta w <0.00005$ was found to be too restrictive, and the solution diverged after  $10^6/2\Omega$.  A detuning criterion of $\delta w < 0.0001$ excludes the coupling with the third lowest parametric instability to the r-mode and, although the evolution remains stable for $>10^9/2\Omega$,  the dynamics differ from evolutions with less restrictive detuning criteria. The absence of the strong damping due to the coupling to the triplet with  third lowest parametric instability threshold allows the mode to grow to larger amplitudes before decaying. Characteristics such as the average energy contained in the system and average r-mode amplitude are roughly the same as observed using a less restrictive detuning criterion. 

\begin{figure}[h]
\includegraphics[width=0.7\columnwidth]{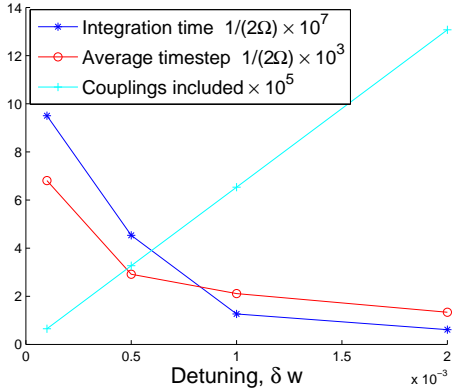}
\caption{(Color online) Comparison of the integration times achieved, average timestep sizes and number of couplings included for various detuning criteria. All evolutions were run on 16 processors of the NCSA Platinum cluster for a period of 24 hours. The temperature was chosen as $T=5\times 10^7$K.} 
\label{d2Comp} 
\end{figure}

The benefit of selecting a strict detuning criterion is displayed in Figure~\ref{d2Comp}.
Not only does the number of couplings scale linearly with detuning, but the average timestep also increases as the maximum allowed $\delta w$ becomes smaller. 
Note that because the runs for large detuning have not yet reached a regime in which many modes become excited, the average stepsizes shown in the Figure~\ref{d2Comp} for $\delta w = 0.001,\ 0.002$ are  overestimates, and can decrease by up to a factor of 100 as more modes start interacting. All subsequent runs reported here are done for a maximum detuning  $\delta w = 0.002$. Although the detailed modal evolutions may differ from that of the complete system, we believe  that the averaged dynamics will not be sensitive to inceases in the maximum $\delta w$. Since many of the couplings are near resonance, the system  displays a sensitive dependence on initial conditions, especially later in the evolution when higher order modes become excited.  Small changes in the round-off accuracy and architecture of different computers results in  modal evolutions that differ in detail (for example, the  positions of sawtooth meanderings of the r-mode in Figures~\ref{Evo5e7}~and~\ref{Evo5e8} shift). Global properties such as the total energy of the system and time averaged modal amplitudes are, however, found to be robust.  

\subsection{Weakly damped evolutions}
 As the temperature increases, the shear damping rates decrease, and the amplitude decay from the initial noise spectrum slows.  The energy from the unstable r-mode has more time to drain into the other modes before they are damped, and therefore numerous modes become important dynamically.      
\begin{figure}[h]
\includegraphics[width=\columnwidth]{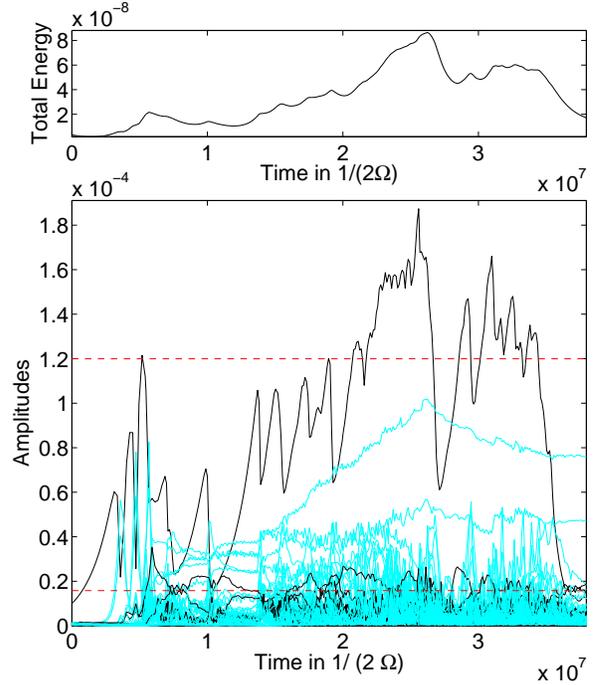}
\caption{Evolution of the $T = 5\times 10^7$K system with $\delta w < 0.002$. The parametric instability thresholds are indicated by horizontal dashed lines} 
\label{Evo5e7} 
\end{figure}
Figure~\ref{Evo5e7} displays an intermediate case,  $T = 5\times 10^7$K,  for which  the  ratios of the decay rates of the daughter modes (414 and 538) involved in the first parametric instability to the growth rate of the  r-mode are  0.14 and 0.18; for the second parametric instability (daughter modes 494 and 592) the ratios are 0.16 and 0.18. Crudely, we would then expect to need the r-mode to couple to about 10 modes with similar damping rates to curb its growth directly.   Instead, energy is rapidly distributed among the daughter modes, which excite many more modes in turn. If one makes a plot like Figure~\ref{modalSpectrum} for the simulation shown in Figure~\ref{Evo5e7}, then it looks very similar to the strongly damped case except that all the amplitudes are displaced upward, so that on average they lie at the initial noise level. 
 Approximately half of the modes in the system  rise above the noise, while only the most strongly damped modes  decay below the noise level. This causes the energy to be shared among the accessible modes. In comparison with a  more weakly damped case,  $T = 5\times 10^8$K displayed in Figure~\ref{Evo5e8}, larger mode amplitudes are attained in certain  modes in Figure~\ref{Evo5e7} even though the energies of the two systems are comparable.      
\begin{figure}[h]
\includegraphics[width=\columnwidth]{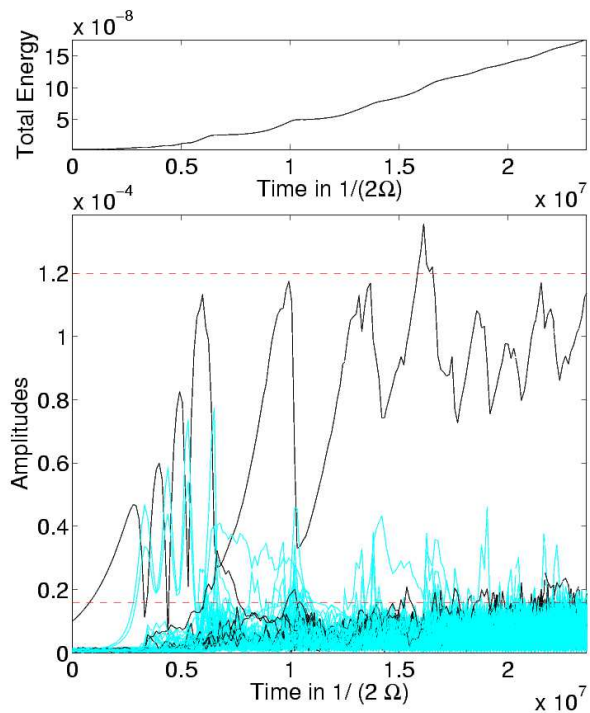}
\caption{Evolution of the $T = 5\times 10^8$K system with $\delta w < 0.002$. The parametric instability thresholds are indicated by horizontal dashed lines. Modal amplitudes are indicated below and the total Hamiltonian energy top.}
\label{Evo5e8} 
\end{figure}

Figure~\ref{Evo5e8} represents the most weakly damped case considered during this investigation. In a realistic neutron star, it lies in the temperature range where the effect of shear viscosity becomes small, and other damping effects such as bulk viscosity begin to dominate. In this slowly damped case, the parametric instability thresholds are very close to their limiting zero damping values. For the first parametric instability the damping rates of the daughters (414 and 538) are 0.0014 and 0.0018  times the r-mode growth rate. 
It would  take approximately 1000 couplings to similarly damped modes of comparable amplitude to curb the r-mode growth.  Our network of 4995 oscillators includes enough modes with sufficient damping to satisfy the criteria for a stable Kolmogorov spectrum \cite{Zak}, namely that the net damping of all the modes is positive. In the weakly damped  regime all the modes rise above the noise. Energy sharing among the daughters is rapid, with few individual daughter modes rising substantially above the average amplitude.  The r-mode, though, maintains an amplitude well above the ever-increasing amplitude level of the daughter modes. Energy tends to be ``trapped'' in the r-mode because it has few direct nearly resonant couplings.
With the exception of the unstable r-mode, the modal spectrum closely resembles the one observed for the Hamiltonian case, Figure~\ref{modalSpectrum}, with a slightly larger  variance in the  modal amplitudes about their mean.  By the end of the simulation no trace of the characteristic shear damping structure could be seen. 

In contrast to the more strongly damped case displayed in Figure~\ref{Evo5e7}, where the total energy peaks and then drops, the total energy in this system has not yet reached a maximum turning point. The  modal amplitudes of  Figure~\ref{Evo5e7} are expected to continue to bounce around  in the same amplitude range as seen in the simulation so far and the energy is not expected to resume monotonic growth. This expectation is supported by simulations with a much more restrictive detuning criterion, $\delta w < 0.0001$  evolved for $2\times 10^8/2\Omega$.  The outcome of the evolution in Figure~\ref{Evo5e8} is less certain. The r-mode might remain at nearly the same amplitude $\sim 10^{-4}$, since any increase in amplitude is severely penalized by an increase in the rate to equipartition, resulting in energy flow out of the r-mode. As can be read off Figure~\ref{HamEntrop}, the time to achieve equipartition for an r-mode amplitude of $1.3\times 10^{-4}$ is $~10^6/2\Omega$ which nicely balances the characteristic driving time due to GR of $1/\gamma_4 \sim 1.9\times 10^6/2\Omega$  . If it were to remain at at this fixed amplitude, the r-mode would pump energy into the sea of inertial modes until they reach  sufficiently high  amplitudes 
that the cumulative damping of the sea of  modes halts the growth in Hamiltonian energy. At this stage, the modal amplitude distribution may change slightly to reflect the stronger damping at the higher order modes.  One of the problems in obtaining this state numerically is that as the energy increases, and the noise level is raised, the nonlinear timescales of the modal oscillations decrease, which cuts the timestep and therefore increases integration times. The current simulation shown in Figure~\ref{Evo5e8} took 20 days on 32 processors of the NCSA Platinum cluster.

\subsection{Rapid Rotation}
\begin{figure}[h]
\includegraphics[width=\columnwidth]{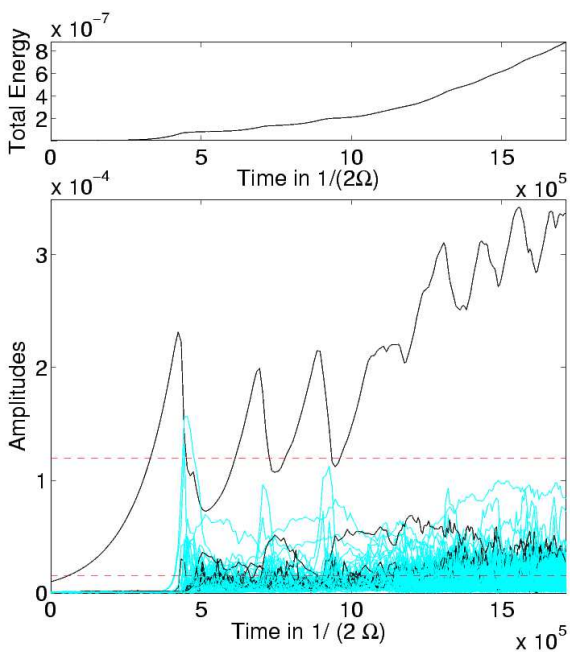}
\caption{Evolution of the $T = 5\times 10^7$K, $e=0.82$ system with $\delta w < 0.002$. The parametric instability thresholds are indicated by horizontal dashed lines. Modal amplitudes are indicated below and the total Hamiltonian energy in the top panel.}
\label{RapidRot} 
\end{figure}

Up until this point the dynamics of a star rotating at $\Omega = 0.37 \sqrt{ \pi G \rho}$, $e=0.5$ have been considered. The maximum rotation rate that can be realistically handled by our perturbative model of Maclaurin spheroids is $\Omega = 0.61 \sqrt{\pi G \rho}$ or $e=0.81267$  where the system displays a bifurcation to the 
Jacobi sequence \cite{Ellips}. To demonstate the most extreme case of rapid rotation we consider the damping for a star  $e=0.82$; this will serve as an upper bound for the maximum possible driving rate that can be represented in our model.  The temperature we use to determine the shear damping is $T=5\times10^7$K (i.e. the intermediate damping case of Figure~\ref{Evo5e7}) . Since the time variable is $2\Omega t$, damping rates become  ``less effective'' for a more rapidly rotating star. The numerical values for the shear damping are effectively decreased by a factor of 0.6; the nonlinear couplings are the same for either $\Omega$ in our units. In these units the driving rate of the most unstable r-mode is $\gamma_4 =7.49\times 10^{-6} (2\Omega)$,  as opposed to  $\gamma_4=5.92 \times 10^{-7} (2\Omega)$ considered previously. This implies that the rate to achieve equipartition will  balance the instability driving rate at an r-mode amplitude  $\sim5\times 10^{-4}$ (see Figure~\ref{HamEntrop}). 
For such a rapidly rotating star, gravitational radiation strongly damps certain low order modes and dominates viscosity in the low order region. However, the number of modes, affected is so small that it does not influence the dynamics at all. The evolution of the modal amplitudes is shown in Figure~\ref{RapidRot}. 
An interesting feature is that the r-mode  grows so rapidly via its linear instabilty that it  passes its first parametric instability threshhold without any permanent slowdown. This is because the instability grows faster than the nonlinear period associated with the first unstable coupling.  (A similar dynamical effect is expected with an artificially ramped up radiation reaction force. If the mode grows too rapidly it bypasses possible instability thresholds and does not excite the daughter modes.)

\section{Discussion}

We have studied the effects of second order nonlinear couplings
among inertial modes of an incompressible star in order to understand 
how they may truncate the CFS instability of the $n=3$, $m=2$
r-mode. Although the system we study is not a realistic version of
a neutron star, it illustrates certain general principles that apply 
to the nonlinear evolution of the instability. 

As in our earlier studies \cite{Katrin1}\cite{Saturation}\cite{JdB3}, 
we have adopted the philosophy that nonlinear effects
can slow or stop the growth of the amplitude of the inertial modes
of a neutron star so that large amplitudes are {\it never} attained.
Thus, we can employ an expansion of the fully nonlinear hydrodynamical
equations of motion, keeping only the lowest order nonlinear terms .
 This philosophy can be justified in retrospect, since as shown in this paper nonlinear effects do indeed limit the dynamics to small amplitudes.

  We may idealize further by restricting attention
only to the inertial mode sector of perturbations of a (slowly)
rotating neutron star. One reason we can do this is that these
modes have frequencies that lie between $\pm 2\Omega$, which are
generally far below the frequencies of the $p$ and $f$ modes of
the star, which have frequencies $\gtrsim\sqrt{\pi G \rho}$ and,
for small mode wavelengths, $\gg \sqrt{ \pi G \rho}$. Thus, we do not
expect any instability involving the r-mode to excite the higher
frequency modes efficiently. Moreover, nonlinear effects are
magnified by near resonances even at small modal amplitudes. Such
resonances are likeliest in the rather dense inertial mode sector.
Indeed, one of the general principles governing the importance
of nonlinearity in weakly nonlinear perturbation theory of the
kind we present is that modes with small relative detunings
dominate the dynamics.

In this paper, we have presented a sequence of calculations of
increasing complexity in order to illustrate the kinds of nonlinear
effects that might arise. We began by treating the simplest system:
three nonlinearly interacting modes. Here, a key concept is parametric
instability: when the amplitude of the unstable ``parent'' mode exceeds 
a well-defined threshold (Equation~\eqref{ParINSTB2}), it will drain energy into 
the pair of ``daughters'' to which it couples. If the daughters
dissipate energy very strongly, then the amplitude of the
unstable parent generally cannot grow more than a factor of a
few beyond its parametric instability threshold. However, even
in such strongly damped cases, the dynamics can be quite complex,
as is shown in Figures \ref{ThreeModSol}(a)-\ref{ThreeModSol}(c). Characteristically, the amplitude
of the parent exhibits a sawtooth evolution and the system as
a whole shows multiply periodic temporal behavior. As the dissipation
drops, though, the three mode system evolves toward a single
period, divergent solution, in which the amplitude of the unstable
mode increases secularly. In a realistic system, additional
modes would be excited at such low dissipation rates. Indeed,
adding the triplet interactions with the second and third 
lowest parametric instability thresholds slowed and even halted
the rate of growth of the instability (Figure \ref{GrowthRates}).

In order to understand what happens in a system involving a
multitude of modes, we began by studying a Hamiltonian system
(i.e. no unstable or damped normal modes). For this system,
we could track the evolution starting from given initial
r-mode amplitudes and noise level. We found that
in general the sea of inertial modes tends toward equipartition, but the behavior
of the r-mode and even a few other lower order modes is
richer. For example, starting with an r-mode amplitude above
its second parametric instability threshold leads to equipartition
in the sea of modes, but the r-mode itself appears to asymptote
toward its first parametric instability amplitude.

We also studied the approach to equipartition as a function of
the initial r-mode amplitude at fixed initial noise level. To measure the 
rate of evolution toward equipartition we
introduced the time-dependent entropy (equation~\eqref{entropyeq} ). In general,
the approach to equipartition is faster for larger initial
r-mode amplitudes. At smaller amplitudes, the approach to
equipartition can become very slow (see Figure~\ref{HamEntrop} ) and may not occur at all. 
At small amplitudes the r-mode, as well as a few
other low order modes, continue to exhibit phase coherent
oscillations up to the end of our simulation at $t\sim 10^7/2\Omega$. 
This maintenance of phase coherence 
signals a failure of the system to become ergodic, and also
warns against assuming the random phase approximation 
too casually in trying to understand nonlinear truncation
of the instability.

Finally, we turned to multimode non-Hamiltonian systems
with growth due to the CFS instability, and damping due
to (shear) viscous dissipation. There are a number of
competing elements that determine how such a complicated
system evolves. The two simplest are the growth or
decay rates of the modes dictated by linear perturbation
theory, but nonlinear interactions introduce several other
important timescales. One was already evident in our study
of the three mode system: in the absence of dissipation,
the modal amplitudes evolve periodically as long as the
total energy is not too large. A second was indicated by
our study of the multimode Hamiltonian system: the rate of
approach to equipartition represents a kind of multi-mode
dissipation rate for the phase coherent low lying modes. Here we
interpret dissipation broadly as the loss of energy
from large scale coherent motion to small scale incoherent
excitations.

For the full multimode system, we found that at strong
damping only a few modes matter to the dynamics. The
saw-tooth, chaotic behavior seen in the three mode system
is also evident in strongly damped multimode systems. Many
of the modes in the sea simply decay via viscous damping,
never having achieved substantial excitation.

At very weak damping, though, numerous modes become
involved in the evolution, which comes to resemble
more closely what we observed in the approach to equipartition
for a Hamiltonian system.  Modal spectra resemble
that of the equipartition, Hamiltonian system, except with larger
variances of some amplitudes about the mean. In the weak
damping regime, the total energy of the system continued to
rise throughout our integration time, but the r-mode amplitude
appeared to level off. At its apparent asymptotic value, the
rate of approach to equipartition -- as indicated by Fig. 6
for our Hamiltonian system -- would be just comparable to
the linear growth rate of the CFS instability.  Thus, in this 
regime, the truncation of the r-mode amplitude itself is due to the
``collective dissipation'' in which energy disappears into
a sea of coupling modes, a completely nonlinear effect. We expect
that if we were to continue the calculations further in
time, the energy growth would continue until the
amplitudes of small scale, more highly damped modes became
large enough for them to dissipate energy as quickly as the
instability causes it to grow.

The behavior outlined above for the weakly damped system also
depends on the relative rates of the CFS instability, approach
to equipartion, and viscous dissipation. We examined this by
increasing the stellar rotation rate, thus increasing the
gravitational radiation rate. For this case, we did not observe
any leveling of the r-mode amplitude, although we also did not
run the simulation long enough (for practical reasons) for
the amplitude to reach the value at which the growth rate
of the CFS instability would be balanced by the rate of
approach to equipartition. Although this does not
indicate any inconsistency in the picture we have presented
for how the instability is limited by nonlinear effects, it
also cautions us that any calculation that enhances the
instability growth rate could overshoot the amplitude of
the r-mode at which it would level off in reality.

Finally, we note that the saturation amplitudes found here 
are substantially lower than those estimated in the cascade
picture in Arras et al. \cite{Saturation}, but are larger than but perhaps comparable to those 
relevant to the spin evolution of a neutron star estimated
by Wagoner \cite{Wagoner}.     
Our study of nonlinear effects
on the development of the r-mode instability also suggests that
the temporal evolution of the modal amplitude is not smooth
on timescales $\sim$ millions of stellar rotation periods.
The erratic behavior seen in our simulations could, if 
realized in nature, 
complicate searches for gravitational
radiation emission that depend on simple, phase coherent
oscillations of the star.

Because of practical limitations our modal expansion is truncated at $n=30$.  Within this truncation the r-mode is found to go unstable first to  modes with $13\leq n\leq 15$. It is possible that some of the dynamics observed in this paper are dependent on our modal truncation.  The most dramatic effect would result if the minimum parametric instablity threshhold were lowered.  In this case the r-mode may go unstable to a very high order mode and dynamics such as equipartition rates may change considerably. It should be noted that this would imply that the maximum amplitude reached is smaller than or equal to those observed in this paper.
Should the mode ever grow to amplitudes comparable to those observed here, all mechanisms explored in the current truncation will still limit its growth.  Our truncated expansion thus sets an upper bound for outcomes with larger $n$. The fact that no lower parametric threshholds were found in the region with $30\leq n <15$ could, however, be an indication that we have gone far enough. Increasing the number of modes will have a secondary effect of increasing the ``specific heat'' or the system since more degrees of freedom have been introduced. This may be particularly relevant to the evolution in  Figure~\ref{Evo5e8}, causing the noise level to rise more slowly.  In a large enough system, the flat spectrum observed in  Figure~\ref{Evo5e8} should have a tail that is grounded below the noise level,  indicating a damping dominated region. 

For more strongly damped regimes the modal dependence of the damping will play a significant role in determining the accessible phase space.  Sources of damping that are on average  weak, but have a few very strongly damped modes will have very little effect on r-mode dynamics; they will only remove a few degrees of  freedom from a very large phase space. On the other hand damping sources that have a fixed scaling, with large regions of modes damped roughly equally, will have a larger effect in  restricting the phase space.  Thus it is  not the maximum damping rate that is important to the dynamics, but rather the distribution of the most weakly damped modes.

For realistic neutron stars the eigenfunction structure will change somewhat and this will affect the couplings to some extent. Changes in the eigenfrequencies could have a more important effect, since subtle changes could shift various daughter modes into or out of close resonance. Such changes could strongly  influence the parametric instability threshholds.

 The modal frequency structure, namely that for each $n$ and $m= 0 \cdots n-1$ there exist  $n-m$ frequencies distributed between $-2 \Omega$ and $2 \Omega$, should persist for more realistic models.  It is this structure that ensures the near resonances among the daughter modes which in turn results in an equipartitioned solution. The rapid increase in equipartition rate with increasing amplitude is also likely to be the result  of the topology of the frequency spectrum and should persist in other models.

Thanks to Larry Kidder for helping with parallelization of the code and numerous other technical details.  
   
This research is supported in part by NSF grants AST-0307273, PHY-9900672 and PHY-0312072 at Cornell University.

\appendix

\section{Three Mode Hamiltonian system}
\label{AppendThreeHams}
Consider the amplitude equations without driving or damping:
\begin{eqnarray}
\dot{c}_\alpha &=& iw_\alpha c_\alpha - 2i w_\alpha\kappa c_\beta c_\gamma \nonumber\\
\dot{c_\beta} &=& iw_\beta c_\beta -  2iw_\beta\kappa c_\alpha c_\gamma^* \nonumber\\
\dot{c_\gamma} &=& iw_\gamma c_\gamma - 2i w_\gamma\kappa c_\alpha c_\beta^* 
\end{eqnarray}

Using amplitude and phase variables $c_i = \sqrt{w_iA_i}e^{i\phi_i}$, $\phi = \phi_\alpha -\phi_\beta-\phi_\gamma$, $\delta\omega = w_\alpha-w_\beta-w_\gamma $ the system can be rewritten as:
\begin{align}
\dot{A_i} &=-  4 \kappa \epsilon_i  \sqrt{w_\alpha w_\beta w_\gamma A_\alpha A_\beta
A_\gamma} \sin \phi \notag\\
\dot{\phi} &=\delta \omega - \frac{2\kappa  w_\alpha w_\beta w_\gamma \cos \phi}{ \sqrt{w_\alpha w_\beta
w_\gamma A_\alpha A_\beta A_\gamma}} \left(- A_\alpha A_\gamma -  A_\alpha A_\beta +  A_\gamma A_\beta\right)
\label{ampphase} 
\end{align}
where $\epsilon_\beta=\epsilon_\gamma = -1$ and $\epsilon_\alpha =1$.
The Hamiltonian for the three mode system with one coupling is \begin{equation}H = L - 4\kappa \cos \phi\ c_\alpha c_\beta c_\gamma \end{equation} 
where  $L =  |c_\alpha|^2 +|c_\beta|^2+|c_\gamma|^2$ denotes the linear energy.

Define $w^3 = w_\alpha w_\beta w_\gamma$ and $\delta \hat{w} = \delta w/w$. Note that  dividing the amplitude equations  \eqref{ampphase} by each other yields \mbox{$A_{\alpha} - A_{\alpha}^0 = \epsilon_i(A_i - A_i^0)$ }, where the superscript $0$ indicates initial value at time $t=0$.    The difference $l=L-L^0$ between the initial linear energy, $L^0$, and the linear energy at time $t$ can be used to rewrite the three variables $A_i$ in terms of the single variable $l$ and the three constants dependent on the initial amplitudes $A_i^0$: 
\begin{equation} A_i = \frac{\epsilon_i}{\delta w} l +A_i^0
\end{equation}
The nonlinear term in the Hamiltonian can now be rewritten as follows
\begin{align}c_\alpha^2 c_\beta^2 c_\gamma^2 =& 
\left(\frac{w}{\delta w}\right)^3 F(l)~,\end{align}
where \begin{align}
F(l) =&(l+\delta w
A_\alpha^0)(l-\delta w
A_\beta^0)(l-\delta w
A_\gamma^0)  \end{align}
The evolution equation for $l$ is  
\begin{equation}
\dot{l} = -4\kappa\delta w c_\alpha c_\beta c_\gamma \sin \phi
\label{eq:ldot}
\end{equation}
or, using the Hamiltonian to rewrite $\sin \phi$ in terms of $c_i$ and thus $l$,
\begin{eqnarray}
\dot{l} &=& \mp \delta w \sqrt{(4\kappa c_\alpha c_\beta c_\gamma)^2
-(L-H)^2}\nonumber\\
&=&  \mp \delta w \sqrt{(4\kappa)^2 \left(\frac{w}{\delta w}\right)^3F(l)
-(l-H_{nl})^2}\label{ldot}
\end{eqnarray}
where $H_{nl} = H-L^0$.

Equation~\eqref{ldot} yields an elliptic integral for the time dependence of $l$ that can be conveniently expressed in terms of Carlson's $R_F$ function \cite{CarlsonCubic}. 
The $R_F$ function is defined by   
\begin{equation} R_F(U_3^2,U_2^2,U_1^2) =\frac{1}{2} \int_0^l\frac{1}{\sqrt{G}} dt \end{equation}
where $G$ is a cubic polynomial:
\begin{align} G(t)=& (t+a_1)(t+a_2)(t+a_3) \notag\\
=& t^3+at^2+bt+c \end{align} 
The quantities $U_i(x)$ can be written in terms of the roots or coefficients of the polynomial:
 \begin{align}
U_i^2 =& U_0^2 + a_i\notag\\
U_0^2 =& \frac{2\sqrt{cG(x)}+2c + bx}{x^2}\notag\\ 
\end{align}
Specializing to our problem,
let \begin{equation}G(l) = F(l) - Y (l-H_{nl})^2\end{equation} where \begin{equation}Y =\frac{1}{(4\kappa)^{2}} \left(\frac{\delta w}{w}\right)^{3}~.\end{equation}   
Thus
\begin{multline} 
G = l^3 + l^2 \left[\delta w(A_\alpha^0-A_\beta^0-A_\gamma^0)-Y\right]\\ + l \left[2H_{nl}Y
+\delta w^2 ( A^0_\gamma A^0_\beta - A^0_\alpha A^0_\beta - A^0_\alpha
A^0_\gamma)\right]\\ + \delta w^3 A^0_\alpha A^0_\beta A^0_\gamma -Y H_{nl}^2
\end{multline}
The time dependence can be found by integrating the equation \eqref{ldot} from $t=0$ to $t$
to yield
\begin{eqnarray}
t& =& \frac{\sqrt{Y}}{\delta w}\int_0^l \frac{1}{\sqrt{G(t)}} dt\nonumber\\ 
&=& \frac{2\sqrt{Y}}{\delta w} R_F(U_3^2,U_2^2,U_1^2) \label{tofl}
\end{eqnarray}
where $U_i^2 = U_0^2 + l_i$. 
Equation~\eqref{tofl} gives $t$ a function of $l$, and can be inverted in terms of the Jacobian elliptic functions.

The phase dependence of $\phi$ can be found using 
\begin{equation}
\sin^2\phi = \frac{G(l)}{F(l)}
\end{equation}
or
\begin{equation}
\cos \phi =-\sqrt{Y} \frac{(H_{nl}-l)}{\sqrt{F(l)}}
\end{equation}

\subsection{Period of Oscillation}

Theoretically the period of oscillation can be obtained directly from equation~\eqref{tofl}. However numerical calculation of the solution turns out to be tricky for a large range of energies and initial values. 
 Scalings of the variables and parameters turns out to be important to ensure accurate results in some cases.

Introduce the parameter  
\begin{equation}X =
\frac{4^2\kappa^2 w_\alpha w_\beta w_\gamma}{\delta w^2 }~.
\end{equation} 
The quantity $1/X$ is the limit of the parametric instability threshold in the absence of damping. Then equation\eqref{ldot} can be rewritten
\begin{multline}
\dot{l_H}
= \pm \sqrt{\delta w X H}  \left[\left(-l_H+\frac{\delta w}{H} \frac{c_\alpha(0)^2}{w_\alpha}\right)\left(l_H+\frac{\delta w}{H}  \frac{c_\beta(0)^2}{w_\beta}\right) \right.\\
\left.\left(l_H+\frac{\delta w}{H} \frac{c_\gamma(0)^2}{w_\gamma}\right)
-\frac{\delta w}{HX} \left(l_H+\frac{H_{nl}}{H}\right)^2   \right]^{1/2} \label{lhdot}
\end{multline}
where $l_H =( L_0-L)/H$. 

Note that the Hamiltonian rescales the detuning, so the smaller the relevant energies the more important the detuning becomes.
Expanding the polynomial in $l$ in the square brackets  in equation~\eqref{lhdot} gives 
\begin{multline}
- l_H^3 + l_H^2\frac{\delta w}{H}\left(D_\alpha - D_\beta - D_\gamma - \frac{1}{X}\right)\\ +
l_H\frac{dw^2}{H^2}\left(D_\alpha D_\beta  + D_\alpha D_\gamma - D_\beta D_\gamma - \frac{2H_{nl}}{X\delta w}\right) \\ +
D_\alpha D_\beta D_\gamma \frac{\delta w^3}{H^3}  
 - \frac{\delta w H_{nl}^2}{H^3X}
\end{multline}
where $D_i = c_i(0)^2/w_i$

The roots of a cubic polynomial of the form $x^3+ax^2+bx+c=0$ can be computed
using the intermediaries \cite{NR}
\[ Q = \frac{a^2-3b}{9} \ \ \ \mbox{and} \ \ \ R = \frac{2a^3-9ab+27c}{54}\]
If $R^2<Q^3$ there are three real roots, which is the only case that will be
considered here. Setting $\theta = \arccos(R/\sqrt{Q^3})$ gives the roots as 
\begin{equation}
x_k = -2\sqrt{Q} \cos (\frac{\theta+2\pi k}{3})-\frac{a}{3}, \ \ k=-1,0,1
\end{equation}    

If the roots of the polynomial in $l_H$ are ordered as $l_1< l_2 < l_3$, and assuming
that 
 the $l=0$ initial conditions lie between roots $l_1$ and $l_2$, the period of
the oscillation $P$ is  
 \begin{eqnarray}
P&=&2 \int_{l_1}^{l_2} \frac{1}{\dot{l_H}} dl_H\nonumber\\
&=& \frac{2}{\sqrt{|\delta w X H|}} \times \frac{2}{\sqrt{l_3-l_1}}
K(k) \nonumber\\
&=&\frac{4}{\sqrt{|\delta w X H|}}R_F(0,l_3-l_2,l_3-l_1)
\end{eqnarray} 
where $k^2 = (l_2-l_1)/(l_3-l_1)$ and 
$ K(k)$ is the
complete elliptic integral of the 1st kind.

Note that for initial conditions with a large parent and two small daughter amplitudes, $k$ gets very
close to one (since $l_3 \rightarrow l_2$) and the integral diverges
logarithmically.

\section{Damped Three Mode System}
\label{parametricAPPENDIX}
This appendix summarizes the main properties of the damped three-mode system relevant to our problem. The various ways of rewriting the equations provide different possible approaches to the numerical integration of the oscillator net. The scheme presented by equation~\eqref{eiwt3mode} has turned out to be particularly fruitful, especially if only modes of small detuning are kept.
Consider the amplitude equations.
\begin{eqnarray}
\dot{c}_\alpha &=& iw_\alpha c_\alpha + \gamma_{GR} c_\alpha - 2i w_\alpha\kappa c_\beta c_\gamma \nonumber\\
\dot{c_\beta} &=& iw_\beta c_\beta - \gamma_{\beta} c_\beta -  2iw_\beta\kappa c_\alpha c_\gamma^* \nonumber\\
\dot{c_\gamma} &=& iw_\gamma c_\gamma - \gamma_{\gamma} c_\gamma - 2i w_\gamma\kappa c_\alpha c_\beta^* 
\end{eqnarray}
Explicitly following the rapid oscilations of the individual oscillators can be avoided by changing variables to $C_j = c_j e^{-iw_jt}$, in which case the equations become
  \begin{eqnarray}
\dot{C}_\alpha &=& \gamma_{GR} C_\alpha - 2i w_\alpha\kappa C_\beta C_\gamma e^{-i\delta w t}\nonumber\\
\dot{C_\beta} &=& - \gamma_{\beta} C_\beta -  2iw_\beta\kappa C_\alpha C_\gamma^* e^{i\delta w t}\nonumber\\
\dot{C_\gamma} &=& - \gamma_{\gamma} C_\gamma - 2i w_\gamma\kappa C_\alpha C_\beta^*e^{i\delta wt} \label{eiwt3mode}
\end{eqnarray}
Using amplitude and phase variables $c_i = |c_i|e^{i\phi_i}$, \mbox{$\phi = \phi_\alpha -\phi_\beta-\phi_\gamma$}, $\delta\omega = w_\alpha-w_\beta-w_\gamma $ gives
\begin{align}
\dot{|c_\alpha|} &=  \gamma_{GR} |c_\alpha| - 2 w_\alpha\kappa |c_\beta| |c_\gamma| \sin(\phi) \notag\\
\dot{|c_\beta|} &= - \gamma_{\beta} |c_\beta |+  2w_\beta\kappa |c_\alpha| |c_\gamma| \sin(\phi) \notag\\
\dot{|c_\gamma|} &=  - \gamma_{\gamma} |c_\gamma| + 2 w_\gamma\kappa |c_\alpha| |c_\beta|\sin(\phi) \notag\\
\dot{\phi} &=  \delta \omega - 2\kappa\cos(\phi)\notag\\
&\left(-\omega_\beta \frac{|c_\alpha| |c_\gamma|}{|c_\beta|} -  \omega_\gamma \frac{|c_\alpha| |c_\beta|}{|c_\gamma|} + \omega_\alpha \frac{|c_\gamma| |c_\beta|}{|c_\alpha|}\right) 
\end{align}
from which the  
stationary solution 
\begin{align}
\tan\phi =&\frac{\gamma_\beta+\gamma_\gamma-\gamma_{GR}}{\delta w} \notag\\
|c_\gamma|^2 =& \frac{\gamma_{GR}\gamma_\beta}{4w_\alpha w_\beta \kappa^2}\left(1+\frac{1}{\tan^2\phi}\right) \notag\\
|c_\alpha|^2 =& \frac{\gamma_{\gamma}\gamma_\beta}{4w_\gamma w_\beta \kappa^2}\left(1+\frac{1}{\tan^2\phi}\right)
\end{align} can be obtained.

\subsection{Parametric instability}

If we treat the parent or driven mode as a known function of time, then in the limit of slow driving the parent can be considered constant \cite{Dziem}. Let $C_i = S_ie^{i\delta w t/2} $
so that equations (\ref{eiwt3mode}) become
  \begin{eqnarray}
C_\alpha &=&\mbox{ constant }\nonumber\\
\dot{S_\beta} &=& (-\frac{i\delta w}{2} - \gamma_{\beta}) S_\beta -  2iw_\beta\kappa C_\alpha S_\gamma^* \nonumber\\
\dot{S_\gamma^*} &=&2i w_\gamma\kappa C_\alpha^* S_\beta  +(\frac{i\delta w}{2} - \gamma_{\gamma}) S^*_\gamma 
\label{paraeqconst}\end{eqnarray}
Let $\bj{S} = [S_\beta \ \ S^*_\gamma]$. Then equations~\eqref{paraeqconst}  becomes $\bj{\dot{S}}=A \bj{S} $, where $A$ is a matrix with trace \mbox{$ T_A = -(\gamma_\beta+\gamma_\gamma)$} and determinant 
\begin{align}d_A = \gamma_\beta \gamma_\gamma + (\delta w/2)^2  -4 w_\beta w_\gamma \kappa^2|C_\alpha|^2+(\gamma_\gamma-\gamma_\beta)i \delta w/2~.\end{align}

The eigenvalues of A are $\lambda = \frac{1}{2}(T_A \pm \sqrt{T_A^2-4d_A})$ and the solutions are stable iff $\Re (\lambda) <0$ where $\Re$ indicates the real part of a complex number. The solution becomes unstable when  
$\Re (\lambda) =0$ or $T_A^2 = [\Re(\sqrt{T_A^2-4d_A)}]^2$  if $T_A <0 $. If  $T_A >0$ the solution is always unstable. 

If a complex number $ u+iv = (x+iy)^2$, then  $u = x^2-y^2,\  v = 2xy$ or $4x^4-v^2=4x^2u$. In our problem $u = T_A^2 - 4 \gamma_\beta \gamma_\gamma -\delta w^2  +16 w_\beta w_\gamma \kappa^2|C_\alpha|^2$, $v= -2(\gamma_\gamma-\gamma_\beta)\delta w$ and $x=T_A$ at the instability point.
As a result
$4T_A^2(- 4 \gamma_\beta \gamma_\gamma -\delta w^2  +16 w_\beta w_\gamma \kappa^2|C_\alpha|^2)=-4 (\gamma_\gamma-\gamma_\beta)^2\delta w^2 $  
and the threshhold value at which the daughters or damped modes become unstable, known as the parametric threshold is  
\begin{equation}|C_\alpha|^2 = |c_\alpha|^2 = \frac{\gamma_\beta \gamma_\gamma}{4w_\gamma w_\beta \kappa^2} \left[1+\left(\frac{\delta w}{\gamma_\gamma+\gamma_\beta}\right)^2\right]
\label{ParINSTB}
\end{equation}      
 The signs of the frequencies are important.  If in the above example $w_\gamma w_\beta < 0$, no parametric instability takes place and the zero solution remains stable for all amplitudes $C_\alpha$. Suppose now that $c_\beta$ is driven and the other two modes are damped for the same coupling.  The instability criterion becomes \begin{equation}|c_\beta|^2 = \frac{\gamma_\alpha \gamma_\gamma}{-4w_\gamma w_\alpha \kappa^2} \left[1+\left(\frac{\delta w}{\gamma_\gamma+\gamma_\alpha}\right)^2\right]\end{equation} and occurs only if $  w_\gamma w_\alpha < 0$
\bibliographystyle{apsrev}
\bibliography{../Paper1}

\end{document}